\documentclass[runningheads]{llncs}

\usepackage{amsmath}
\usepackage{tabularx} 
\usepackage{multirow} 
\usepackage{caption} 
\usepackage{booktabs} 
\usepackage{adjustbox} 
\usepackage{subcaption}
\usepackage{amsmath}
\usepackage{epsfig}
\usepackage{makecell}
\usepackage[T1]{fontenc}
\usepackage{amsfonts}
\usepackage{xcolor}

\usepackage{graphicx}
\usepackage{fontawesome5}

\begin{document}
\title{Robust AI-Synthesized Image Detection via Multi-feature Frequency-aware Learning}
\titlerunning{Robust AI Detection via MFL}
%
%
\author{Hongfei Cai \inst{1}
\and
Chi Liu\inst{1}\faIcon{envelope}
\and
Sheng Shen\inst{2}
\and
Youyang Qu\inst{3,4}
\and
Peng Gui\inst{5}
%
}
\authorrunning{H. Cai et al.}
%
\institute{Faculty of Data Science, City University of Macau, Macao SAR, China 
\and
Design and Creative Technology Vertical, Torrens University Australia, NSW, Australia
\and
Shandong Provincial Key Laboratory of Computer Networks, Ministry of Education, Shandong Computer Science Center, Qilu University of Technology (Shandong Academy of Sciences), Jinan, China
\and
School of Computer Science and Engineering, Wuhan Institute of Technology, Wuhan, China\\
\faIcon{envelope} Corresponding author: \email{chiliu@cityu.edu.mo}
}


%
\maketitle              
\begin{abstract}
The rapid progression of generative AI (GenAI) technologies has heightened concerns regarding the misuse of AI-generated imagery. To address this issue, robust detection methods have emerged as particularly compelling, especially in challenging conditions where the targeted GenAI models are out-of-distribution or the generated images have been subjected to perturbations during transmission. This paper introduces a multi-feature fusion framework designed to enhance spatial forensic feature representations with incorporating three complementary components, namely noise correlation analysis, image gradient information, and pretrained vision encoder knowledge, using a cross-source attention mechanism.  Furthermore, to identify spectral abnormality in synthetic images, we propose a frequency-aware architecture that employs the Frequency-Adaptive Dilated Convolution, enabling the joint modeling of spatial and spectral features while maintaining low computational complexity. Our framework exhibits exceptional generalization performance across fourteen diverse GenAI systems, including text-to-image diffusion models, autoregressive approaches, and post-processed deepfake methods. Notably, it achieves significantly higher mean accuracy in cross-model detection tasks when compared to existing state-of-the-art techniques. Additionally, the proposed method demonstrates resilience against various types of real-world image noise perturbations such as compression and blurring. Extensive ablation studies further corroborate the synergistic benefits of fusing multi-model forensic features with frequency-aware learning, underscoring the efficacy of our approach.

\end{abstract}
\section{Introduction}
\label{sec:intro}
The recent progress in generative AI (GenAI) models, such as Generative Adversarial Networks (GANs) and Diffusion models, empowers adversaries to fabricate realistic images at low cost. This poses a significant threat to digital integrity and authenticity. As a response, the detection of AI-synthesized images has become increasingly crucial as an initial safeguard for AI-generated content.

There are two persistent challenges remaining in AI-synthesized image detection: the generalization ability and the robustness of detectors. As the GenAI models behind fake image generation continue evolving, a detector trained on specific GenAI models should be able to generalize to identify previously unseen GenAI models. Furthermore, considering the prevalence of AI-generated images distributed online, which may be subject to various transmission noises such as compression and blur, it is essential for a trained detector to maintain robustness against typical image perturbations.

Recently, various AI-synthesized image detectors have been proposed, utilizing forensic features from both the image domain, like texture details   \cite{10.1007/978-3-030-58574-7_7}, noise relationships   \cite{Tan_2024_CVPR}, and gradients   \cite{Tan_2023_CVPR}, and the frequency domain, such as spectral distribution   \cite{pmlr-v119-frank20a}, Fourier amplitude   \cite{9035107}, and DCT coefficients   \cite{pmlr-v119-frank20a}. Additionally, some methods employ feature fusion to integrate these diverse sources of features   \cite{10286083}. However, there still remains a gap in addressing the cross-model generalization and noise robustness problems. Methods that rely on a single feature source are vulnerable to changes in target GenAI models, particularly when the architecture of test GenAI models shifts, e.g., from GANs to diffusion model. Moreover, some features may be typically sensitive to specific noise perturbations; for example, online image compression will significantly compromise the original frequency-domain feature representations   \cite{pmlr-v119-frank20a}. Multi-feature fusion strategies present a promising avenue for addressing the limitations of single-feature representations; however, there persists a requisite for more sophisticated and effective fusion mechanisms that merit further investigation.

To address these challenges, we propose a novel multi-feature fusion and frequency-aware learning framework for generalizable and robust AI-synthesized image detection. Our method first creates a strong forensic multi-feature representation by integrating noise relationship features, image gradient features, and knowledge from pretrained large vision encoders with a cross-source attention module. The noise and gradient features provide reliable forensic clues   \cite{Tan_2023_CVPR,Tan_2024_CVPR}, while the pretrained knowledge, which characterizes the distribution of natural images, enhances the feature representation by sharpening the decision boundary between natural and artificial images   \cite{Ojha_2023_CVPR}. The integrated feature is subsequently fed to a frequency-aware learning backbone. The frequency-aware learning backbone involves incorporating residual learning with the Frequency-Adaptive Dilated Convolution (FADC)   \cite{Chen_2024_CVPR}. This design improves the detector's capability of concurrently capturing both spatial information and frequency information from the previously fused feature, with a particular focus on the local frequency dynamics. Meanwhile, FADC helps reduce network complexity and improves computational efficacy to obtain a lightweight detector. 

To verify the effectiveness of the proposed multi-feature frequency-aware learning framework, we conduct extensive experiments targeting the detection of \emph{fourteen} GenAI models, spanning over GAN-like models, diffusion-like text-to-image models, autoregressive models, low-level and perceptual image processing models, as well as post-rendered deepfakes. We also compare our method with various methods including widely used classifiers such as ResNet and ViT, and some recent state-of-the-art baselines. Ablation studies are additionally provide to demonstrate the rationale of multi-feature fusion. The results confirm that our method achieves exceptional accuracy, generalizability, and robustness in detecting AI-synthesized images. 

To summarize, the contributions of this paper include:
\begin{itemize}
\item We propose a novel framework for AI-synthesized image detection, which significantly improves the cross-model generalization ability and noise robustness of detection.
\item We devise a multi-feature fusion mechanism that enables the model to adaptively incorporate noise relationships, image gradients, and knowledge from pretrained large vision encoders for robust forensic feature representation.  
\item We introduce a frequency-aware learning backbone that effectively integrates global spatial and local frequency information using a lightweight design with low computational cost. 
\item Our extensive experimental evaluation shows that our framework excels in accuracy, generalization, and robustness, surpassing multiple baselines across various test settings.
\end{itemize}

\section{Related Work}
\paragraph{Image-based Detection Methods} focus on analyzing pixel-level and regional features within images to identify potential forgeries. These features include texture details  \cite{10.1007/978-3-030-58574-7_7}, color distribution  \cite{8803740}, saturation  \cite{8803661}, edge information  \cite{9468380}, and other visual cues. For example, some studies have explored the use of gradient information to detect inconsistencies in image textures and edges that are characteristic of AI-generated content  \cite{9468380}. These methods are particularly effective in identifying forgeries where the synthesis process introduces subtle but detectable anomalies in the spatial domain. However, they can be sensitive to image quality and resolution, and may struggle with high-quality forgeries that closely mimic real images. For instance, recent work demonstrated that spatial anomalies can be effectively removed through trace removal attacks, highlighting the limitations of relying solely on spatial features for robust detection  \cite{liu2022making}.

\paragraph{Frequency-based Detection Methods} convert images from the spatial domain to the frequency domain, utilizing features such as high-frequency artifacts to identify synthetic content  \cite{pmlr-v119-frank20a}. These methods leverage the fact that AI-generated images often exhibit distinct frequency patterns, such as periodic structures or anomalies in the high-frequency components, which are not typically present in natural images. For instance, the use of Fourier transforms or wavelet transforms allows for the extraction of frequency features that can reveal the presence of synthetic elements  \cite{pmlr-v119-frank20a}. The frequency domain fingerprinting of GenAI models for task-specific forensics was further explored  \cite{liu2024disentangling}. Frequency-based methods are useful for detecting GenAI models that leave distinct frequency anomaly; however, they may be highly susceptible to image noise or other frequency-domain disturbances  \cite{zhou2022adversarial}. 

\paragraph{Feature-fusion Detection Methods} combine various image features from different layers, such as visual features, frequency features, or multi-modal data, to enhance the accuracy and robustness of detection  \cite{10286083,10246417}. By integrating information from multiple sources, these methods aim to capture both spatial and frequency-domain characteristics of images, thereby improving the detection performance. For example, some studies have proposed frameworks that fuse global and local features, leveraging the complementary nature of different feature sets to achieve better generalization across various forgery types and datasets  \cite{10286083}. Additionally, feature-fusion methods often incorporate advanced techniques such as attention mechanisms to dynamically weigh the importance of different features, further optimizing the detection process  \cite{10246417}. These methods are generally more robust to variations in forgery techniques and image quality, making them suitable for practical applications where diverse forgery scenarios are encountered. A novel multi-view completion representation for robust GAN-generated image detection was proposed, which effectively integrates multi-scale features and cross-view information to enhance detection robustness  \cite{liu2023towards}.

\section{Method}
\begin{figure*}[ht]
    \centering
    \includegraphics[width=\textwidth]{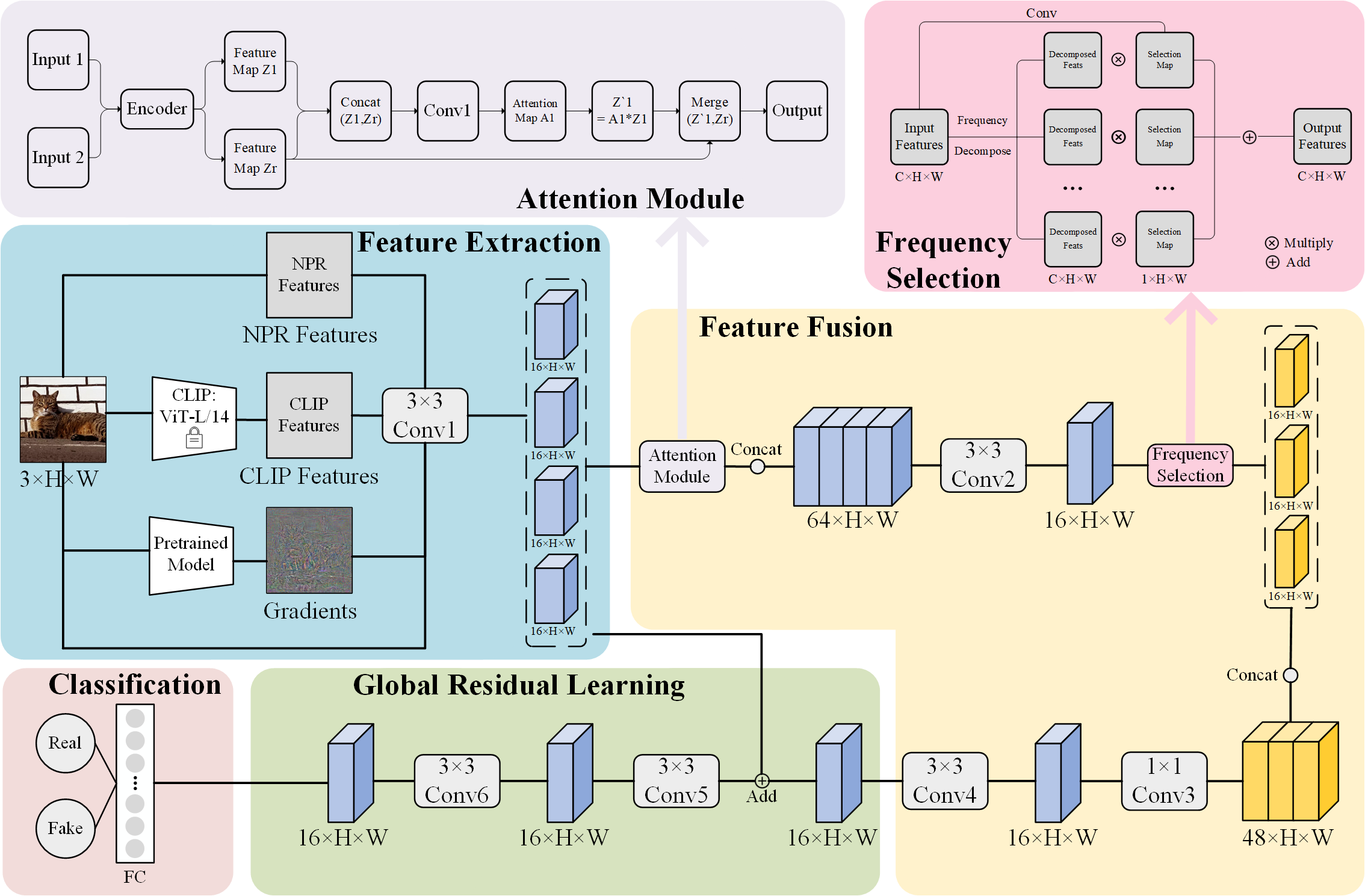}
    \caption{The architecture of the proposed multi-feature frequency-aware learning model. The multi-branch network integrates CLIP semantic features, transformation gradients, and NPR (noise pattern residual) features. Frequency decomposition via Discrete Wavelet Transform separates low- and high-frequency components, followed by residual-enhanced feature refinement and attention-based feature weighting. Final classification is achieved through hierarchical convolutional blocks and fully connected layers.}
    \label{fig:framework}
\end{figure*}

Our framework for detecting AI-synthesized images integrates four complementary feature representations within a frequency-aware architecture, addressing both spatial and spectral forensic patterns. As shown in Fig. \ref{fig:framework}, the design motivation stems from three key observations:(i)Modern generative models exhibit semantic inconsistencies despite visual realism, necessitating global semantic analysis;(ii)Manipulation artifacts manifest as localized spatial discontinuities;(iii)Synthetic images often contain frequency-domain anomalies across sub-bands.Accordingly, the pipeline employs three parallel processing branches followed by frequency-domain refinement. The CLIP-ViT module leverages pretrained vision-language embeddings to capture semantic coherence, motivated by its proven cross-modal alignment capabilities on large-scale datasets. Concurrently, a transformation-based spatial gradient analyzer detects local pixel-level inconsistencies through Sobel operators, targeting common artifacts in generated image boundaries. The third branch computes noise pattern residuals (NPR) via guided filtering to isolate structural anomalies in texture regions.

These heterogeneous features are concatenated and processed through frequency selection modules that apply Discrete Wavelet Transform (DWT), decomposing signals into approximation (low-frequency) and detail (high-frequency) components. This design explicitly models frequency-space interactions, as synthetic images often exhibit abnormal energy distributions in specific sub-bands. Subsequent global residual blocks employ identity mappings to amplify discriminative patterns while mitigating gradient vanishing. A spatial attention mechanism dynamically reweights feature maps based on local artifact severity, focusing computation on suspicious regions. The network architecture utilizes stacked 3×3 convolutions with ReLU activation and batch normalization, progressively abstracting features through hierarchical processing before final binary classification via fully connected layers. This multi-stage design enables joint modeling of complementary forensic cues while maintaining parameter efficiency through modular components.

\subsection{Neighboring Pixel Relationships and Gradient Features}

\subsubsection{Neighboring Pixel Relationships (NPR):}
Inspired by the inherent up-sampling patterns in generative networks \cite{Tan_2024_CVPR}, we explicitly model local pixel correlations through a grid-based difference operator. Given an input image \( I \in \mathbb{R}^{H \times W \times 3} \), we first partition it into \( W \times H \) regular grids \( \{v_c^I\}_{c=1}^{W \times H} \), where each grid contains \( l \times l \) pixels (\( l=2 \) for common \( 2\times \) up-sampling). The NPR feature tensor \( \hat{V}^I \in \mathbb{R}^{W \times H \times (l^2 - 1)} \) is computed through exhaustive pairwise differences within each grid:

\begin{equation}
\hat{v}_c^I = \left\{ w_i - w_j \mid \forall w_i \in v_c^I,\ 1 \leq j \leq l^2 \right\} \quad \forall c \in \{1,\dots,W \times H\}
\end{equation}
where  \( I \) denotes the input image with height \( H \), width \( W \), and 3 color channels. The grids \( \{v_c^I\} \) are partitioned from the image, each containing \( l \times l \) pixels. The resulting NPR feature \( \hat{V}^I \) captures local pixel correlations by computing pairwise differences within each grid \( v_c^I \). The pixel values within the grid are denoted by \( w_i \) and \( w_j \), and \( c \) is the index of the grid. This operation amplifies characteristic grid artifacts from transposed convolutions in GANs, particularly visible in pseudo-periodic textures like hair strands and fabric weaves. The choice of \( l=2 \) corresponds to the most prevalent up-sampling factor in modern generators.

\subsubsection{Gradient Features:}
To capture complementary artifact signatures, we compute gradient maps using a fixed CNN backbone \( M \) (ResNet-50 pretrained on ImageNet). For each input image \( I_i \in \mathbb{R}^{H \times W \times 3} \), we compute the gradient response \( G \in \mathbb{R}^{H \times W \times 3} \) through guided backpropagation:

\begin{equation}
G = \frac{\partial \sum_{k=1}^K M_k(I_i)}{\partial I_i}
\end{equation}
where  \( M \) denotes the fixed CNN backbone (ResNet-50), \( I_i \) is the input image, and \( G \) is the gradient response map. \( M_k \) represents the \( k \)-th channel output of the last convolutional layer, and \( K \) is the number of channels in that layer. The gradient features highlight high-frequency artifacts in diffusion model outputs while preventing overfitting to training data statistics by keeping the CNN parameters frozen. These features are particularly effective for detecting boundary discontinuities in synthetic shadows and reflections.

\subsection{Incorporating Pretrained Semantic Priors} 
We integrate semantic priors from CLIP-ViT-L/14 \cite{Ojha_2023_CVPR} to distinguish authentic natural image statistics. Given an image $x$, we extract the CLIP feature vector $\phi_x \in \mathbb{R}^{768}$ from the final transformer layer before the projection head, preserving both semantic and textural information.

A multi-head cross-attention mechanism dynamically fuses CLIP's global semantics with local artifact features (NPR $\oplus$ Gradients). Let $F_{\text{local}} = [\hat{V}^I \oplus G] \in \mathbb{R}^{L\times C}$ denote the concatenated local features ($L=W\times H$ spatial positions, $C$ channels). The fusion process is formalized as:

\begin{equation}
\begin{aligned}
Q &= W_Q \phi_x \in \mathbb{R}^{d_k} \\
K &= W_K F_{\text{local}} \in \mathbb{R}^{L\times d_k} \\
V &= W_V F_{\text{local}} \in \mathbb{R}^{L\times d_v} \\
\text{Attention}(Q,K,V) &= \text{Softmax}\left(\frac{QK^\top}{\sqrt{d_k}}\right)V \in \mathbb{R}^{d_v}
\end{aligned}
\end{equation}
where $W_Q, W_K, W_V$ are learnable projections. The softmax temperature $\sqrt{d_k}$ stabilizes gradient flow during training. This attention mechanism enables adaptive feature recalibration, where CLIP's semantic vectors (queries) selectively amplify discriminative local artifacts (values) via compatibility scores with keys.

\subsection{Frequency-Aware Residual Learning} 

In our frequency-aware backbone, each Frequency-Adaptive Dilated Convolution (FADC) block is designed to jointly model spatial and spectral information through a multi-path architecture, which includes Frequency Selection to balance high- and low-frequency components. The detailed formulation is as follows:

\paragraph{High-frequency Path}
\begin{equation}
 Y_{\text{FADC}}(p) = \sum_{k=1}^{K^2} \left( w_{k}^{\text{low}} + w_{k}^{\text{high}} \cdot \lambda_{h}(p) \right) \cdot X(p + \Delta p_{k} \times \hat{D}(p))
\end{equation}
where  $Y_{\text{FADC}}(p)$ represents the output feature at spatial position $p$ from the high-frequency path, where $K$ is the kernel size, $w_{k}^{\text{low}}$ and $w_{k}^{\text{high}}$ are the decomposed convolution weights for low-frequency and high-frequency components, respectively. $\lambda_{h}(p)$ is a dynamic weight factor for high-frequency components, and $\hat{D}(p)$ is the adaptive dilation rate.

\paragraph{Low-Frequency Path}
\begin{equation}
Y_{\text{skip}}(p) = X(p)
\end{equation}
This path directly preserves the low-frequency content of the input feature map $X$ through a skip connection.

\paragraph{Frequency Adaptation Module}
\begin{equation}
\hat{D}(p) = \text{ReLU}(f_{\theta}(X(p))) \times D_{\text{base}}
\end{equation}
where  $\hat{D}(p)$ is the adaptive dilation rate dynamically adjusted based on the local frequency content of the input feature. $f_{\theta}$ is a lightweight frequency predictor implemented as a depth-wise convolution layer, and $D_{\text{base}}$ is a predefined base dilation rate.

\paragraph{Frequency Selection Module}
\begin{align}
X_{b} &= \mathcal{F}^{-1}(M_{b} \cdot \mathcal{F}(X)) \\
\hat{X}(p) &= \sum_{b=0}^{B-1} A_{b}(p) \cdot X_{b}(p)
\end{align}
where  $\mathcal{F}$ and $\mathcal{F}^{-1}$ denote the Fourier Transform and Inverse Fourier Transform, respectively. $M_{b}$ is a binary mask to extract the $b$-th frequency band, and $A_{b}(p)$ is a spatially variant reweighting map for the $b$-th frequency band. $\hat{X}(p)$ is the frequency-balanced feature map.

\paragraph{Output Integration}
\begin{equation}
Y_{\text{out}}(p) = \text{ReLU}(Y_{\text{FADC}}(p) + Y_{\text{skip}}(p) + \hat{X}(p))
\end{equation}

The final output $Y_{\text{out}}(p)$ integrates the high-frequency features from the FADC path, the low-frequency features from the skip connection, and the frequency-balanced features from the Frequency Selection module.

\subsection{Optimization Objective} 
The network is trained end-to-end using a class-balanced focal loss:

\begin{equation}
\mathcal{L} = -\frac{1}{N}\sum_{i=1}^N \left[ \alpha y_i(1-p_i)^\gamma \log p_i + (1-\alpha)(1-y_i)p_i^\gamma \log(1-p_i) \right]
\end{equation}
where $p_i = \sigma(\text{logit}_i)$ is the predicted probability, $\alpha = \frac{N_{\text{fake}}}{N_{\text{real}} + N_{\text{fake}}}$ balances class frequencies, and $\gamma=2$ focuses training on hard examples. The temperature parameter $\gamma$ smooths the loss landscape for improved convergence stability.

\section{Experiments}

\subsection{Target Generative Models}

Since new methods of creating fake images are always emerging, training on images from one generative model is standard practice and testing the model's ability to detect fake images from other, unseen models is standard practice. We follow the protocol outlined in \cite{Wang_2020_CVPR}, using real and fake images from ProGAN \cite{DBLP:journals/corr/abs-1710-10196} for training. During evaluation, we consider a diverse set of generative models. Initially, we assess performance on models referenced in \cite{Wang_2020_CVPR}: ProGAN, StyleGAN \cite{Karras_2019_CVPR}, BigGAN \cite{DBLP:journals/corr/abs-1809-11096}, CycleGAN \cite{Zhu_2017_ICCV}, StarGAN \cite{Choi_2018_CVPR}, and GauGAN \cite{Park_2019_CVPR}. Each generative model provides a distinct set of real and fake images for our analysis. Furthermore, we extend our evaluation to include the guided diffusion model \cite{NEURIPS2021_49ad23d1}, trained on the ImageNet dataset, and recent text-to-image generation models: Latent Diffusion Model (LDM) \cite{Rombach_2022_CVPR}, Glide \cite{DBLP:journals/corr/abs-2112-10741}, as well as the autoregressive model DALL-E \cite{Dayma_DALL·E_Mini_2021}.

\subsection{Baseline Detectors}

We compare our approach with several state-of-the-art baselines: (i) A classification network trained on real and fake images from ProGAN using binary cross-entropy loss, as detailed in \cite{Wang_2020_CVPR}. This network utilizes a ResNet-50 pre-trained on ImageNet. We also consider a variant of this approach where the backbone network is changed to CLIP-ViT to align with our feature space methodology. (ii) A patch-level classification method proposed in \cite{10.1007/978-3-030-58574-7_7} that truncates a ResNet or Xception network to focus on smaller receptive fields when classifying images as real or fake. (iii) A classification network trained on co-occurrence matrices of real and fake images, which has been shown to be effective in image steganalysis and forensics \cite{10.1145/3082031.3083247}. (iv) A classification network trained on the frequency spectrum of real and fake images, capturing artifacts present in GAN-generated images \cite{9035107}. (v) The method proposed by \cite{Ojha_2023_CVPR}, which leverages the feature space of CLIP-ViT for nearest neighbor classification (NN) or linear classification (LC). We focus on the NN variant for comparison.

\subsection{Evaluation Metrics and Settings}

Our experimental configuration follows established multimedia forensic protocols \cite{10.1007/978-3-030-58574-7_7,1903.06836,Wang_2020_CVPR,9035107}, evaluating detection performance through classification accuracy (ACC) and average precision (AP) scores. To ensure fair comparison, we calibrate classification thresholds using a held-out validation set containing samples from all target generative models.

The training paradigm employs stochastic gradient descent with momentum $0.9$, initial learning rate $1 \times 10^{-4}$, and batch size $32$, processing $3$-channel $256 \times 256$ RGB images. We implement linear warmup for the first $500$ iterations followed by cosine learning rate decay. Models train for $20$ epochs with cross-entropy loss, using random seed $1$ for deterministic weight initialization. The architecture utilizes Adam optimization ($\beta_1=0.9$, $\beta_2=0.999$) with weight decay $5 \times 10^{-4}$, incorporating batch normalization and dropout ($p=0.2$) after convolutional layers. All experiments run on PyTorch 2.0 with mixed-precision training, evaluated on an NVIDIA RTX 4090 GPU.

\subsection{Generalization of Detection}

The proposed multi-feature frequency-aware learning framework demonstrates excellent generalization capabilities, particularly on out-of-distribution (OOD) generative models, including both GANs and Diffusion models. As shown in Table \ref{ganacc} and Table \ref{dacc}, our method achieves high accuracy across multiple unseen GAN and diffusion models, significantly outperforming several state-of-the-art baselines. This indicates that the framework is highly effective in detecting AI-synthesized images even when trained on a single generative model (ProGAN) and tested on diverse unseen architectures.

\begin{figure}
    \centering
    \includegraphics[width=0.87\linewidth]{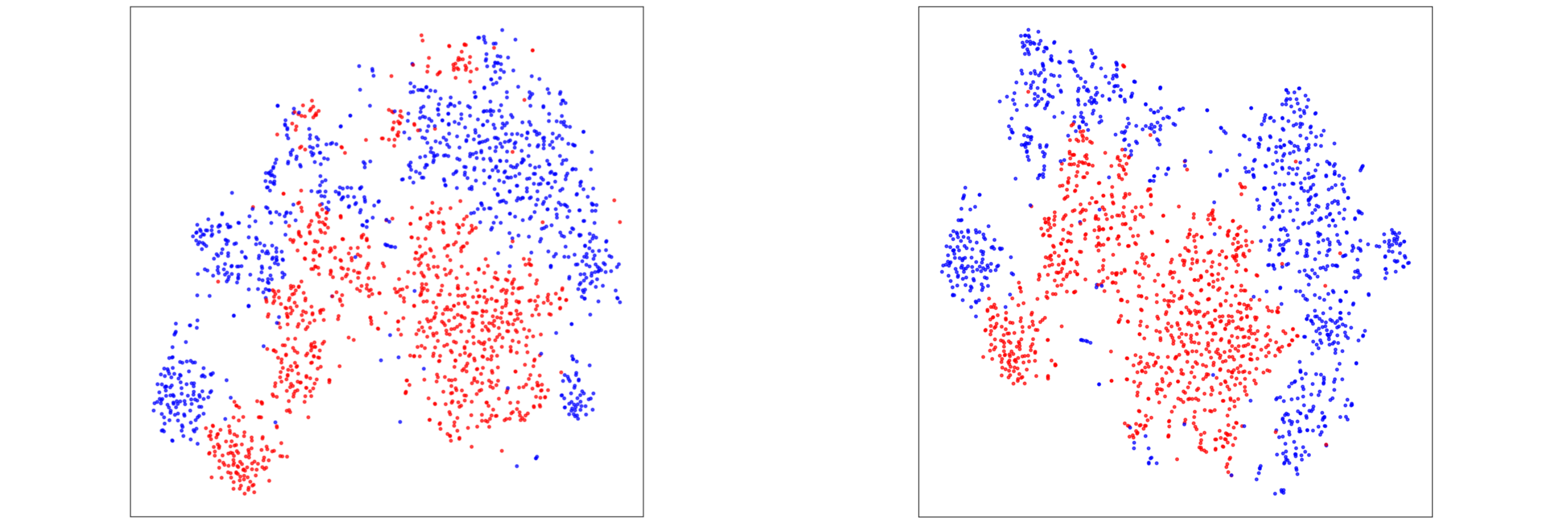}
    \caption{The t-SNE visualization of 2000 test images demonstrates the effectiveness of a frequency domain module in enhancing binary classification accuracy between real (blue) and fake (red) images. The left panel represents the feature distribution without the frequency domain module, while the right panel shows the improved separation after its application.}
    \label{tsne}
\end{figure}

The superior performance on OOD diffusion models is notable, as these models are known for their high visual quality and realism, which often pose challenges for existing detection methods. Our framework's ability to generalize across such diverse and advanced generative models highlights its robustness and adaptability. The effectiveness of our approach can be attributed to the integration of multi-modal features and the frequency-aware learning backbone. These components enable the model to capture both spatial and spectral anomalies that are characteristic of AI-generated images, thereby enhancing the decision boundary between natural and synthetic images. The results in Table \ref{XR} further support this conclusion by demonstrating the importance of each feature branch in achieving robust detection. Additionally, the t-SNE visualization in Fig. \ref{tsne} illustrates the improved separation between real and fake images when incorporating the frequency domain module, confirming the framework's ability to enhance feature discriminability.

\begin{table}[htbp]
  \centering
  \caption{The average accuracy (ACC) of detection across various GAN models. Detectors were trained with real and ProGAN images.}
  \label{ganacc}
  \resizebox{1\textwidth}{!}{
  \setlength{\tabcolsep}{1.1mm}{
  \begin{tabular}{@{}lccccccc@{}} 
    \toprule
    Detection Method & ProGAN & CycleGAN & BigGAN & StyleGAN & GauGAN & StarGAN & Total \\
    \midrule
    CNN Detection \cite{Wang_2020_CVPR} & 98.94  & 78.80  & 60.62  & 60.56  & 66.82  & 62.31  & 71.34  \\
    Patch Classifier \cite{10.1007/978-3-030-58574-7_7} & 94.38  & 67.38  & 64.62  & 82.26  & 57.19  & 80.29  & 74.35  \\
    Co-occurrence \cite{1903.06836} & 97.70  & 63.15  & 53.75  & \textbf{92.50} & 51.10  & 54.70  & 68.82  \\
    Freq-spec \cite{9035107} & 44.90  & \textbf{99.90} & 50.50  & 49.90  & 50.30  & \textbf{99.70} & 65.87  \\
    UniFD \cite{Ojha_2023_CVPR} & 99.54  & 93.49  & 88.63  & 80.75  & \textbf{97.11} & 98.97  & \textbf{93.08} \\
    \textbf{Ours} & \textbf{99.64} & 90.11  & \textbf{92.55} & 87.72  & 94.29  & 93.34  & 92.94  \\
    \bottomrule
  \end{tabular}}}
\end{table}

\begin{table}[htbp]
  \centering
  \scriptsize
  \caption{The average accuracy (ACC) of detecting images generated by various diffusion models. Detectors were trained with real and ProGAN images.}
  \label{dacc}
  \resizebox{1\textwidth}{!}{
  \setlength{\tabcolsep}{1.1mm}{
  \begin{tabular}{@{}lccccccccc@{}} 
    \toprule
    Detection Method & \multicolumn{3}{c}{LDM} & \multicolumn{3}{c}{Glide} & Guided & DALL-E & Total \\
    \cmidrule(r){2-4} \cmidrule(r){5-7}
          & 200 steps & 200w/CFG & 100 steps & 100 27 & 50 27 & 100 10 &       &       &       \\
    \midrule
    CNN Detection \cite{Wang_2020_CVPR} & 50.74 & 51.04 & 50.76 & 52.15 & 53.07 & 52.06 & 50.66 & 53.18 & 51.71 \\
    Patch Classifier \cite{10.1007/978-3-030-58574-7_7} & 79.09 & 76.17 & 79.36 & 67.06 & 68.55 & 68.04 & 65.14 & 69.44 & 71.61 \\
    Co-occurrence \cite{1903.06836} & 70.70 & 70.55 & 71.00 & 70.25 & 69.60 & 69.90 & 60.50 & 67.55 & 68.76 \\
    Freq-spec \cite{9035107} & 50.40 & 50.40 & 50.30 & 51.70 & 51.40 & 50.40 & 50.90 & 50.00 & 50.69 \\
    UniFD \cite{Ojha_2023_CVPR} & 91.29 & 72.02 & 91.29 & 89.05 & 90.67 & 90.08 & 71.06 & 81.47 & 84.62 \\
    \textbf{Ours} & \textbf{94.40} & \textbf{80.60} & \textbf{94.30} & \textbf{91.70} & \textbf{91.10} & \textbf{91.90} & \textbf{95.10} & \textbf{86.90} & \textbf{90.75} \\
    \bottomrule
  \end{tabular}}}
\end{table}

\begin{table}[htbp]
\centering
\caption{The impact of CLIP Features, Gradients, and NPR Features on detection is illustrated by showing the performance changes after feature removal. "$-$" denotes the feature branch has been excluded. "ALL" denotes the performance of the complete model.}
\label{XR}
\resizebox{0.6\textwidth}{!}{
  \setlength{\tabcolsep}{1.7mm}{
\begin{tabular}{lcccc}
\toprule
\textbf{Model} & \textbf{--$f_{Clip}$} & \textbf{--$f_{Grad}$} & \textbf{--$f_{NPR}$} & \textbf{ALL} \\ 
\midrule
GAN models      & 74.23                  & 53.45                  & 62.17                  & \textbf{90.56}           \\
Diffusion models & 91.34                  & 55.67                  & 61.89                  & \textbf{92.78}           \\
\bottomrule
\end{tabular}}}
\end{table}

\subsection{Robustness against Post-Processing Operations}
\begin{figure}[htbp]
    \centering
    \includegraphics[width=0.95\linewidth]{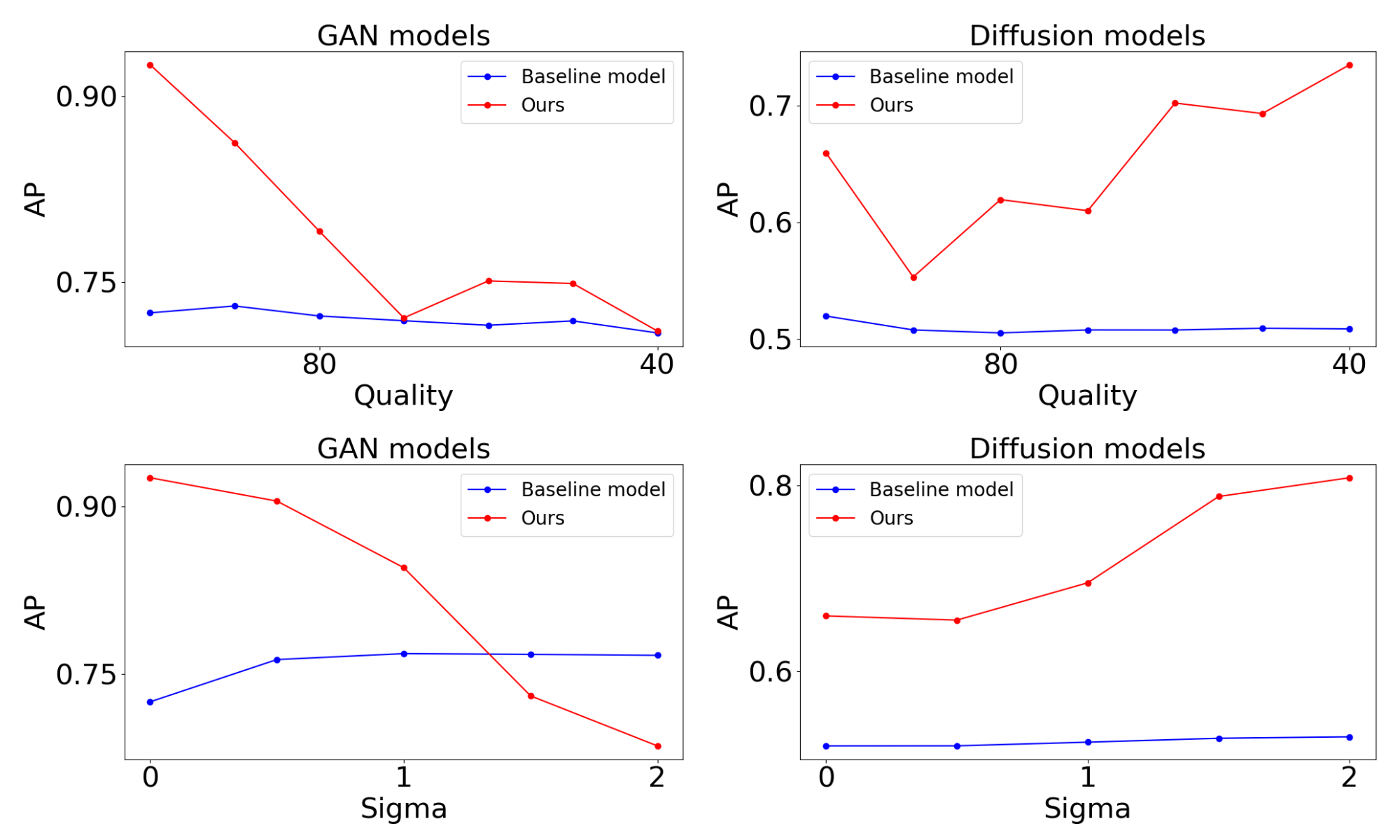}
    \caption{Robustness of different detection methods to various image compressions and noises. Average AP scores of the baseline model  \cite{Wang_2020_CVPR} and ours across two types of generative models (GAN and Diffusion) under different qualities of JPEG compression (top row) and different levels of Gaussian noise (Bottom row) are compared.}
    \label{Robustness}
\end{figure}

To evaluate the robustness of our classifiers against potential evasion tactics, we assess their performance under common image processing operations such as JPEG compression and Gaussian blurring. We select the baseline from  \cite{Wang_2020_CVPR} because it is a widely-adopted strategy for enhancing robustness to image perturbations in AI-generated content detection and offers a well-established benchmark. Figure \ref{Robustness} illustrates the comparison results of our method and the baseline one under varying levels of JPEG compression and Gaussian blurring. Compared to the baseline, our method exhibits higher AP scores, especially in out-of-distribution detections where the detector is trained on ProGAN images while tested on Diffusion images). This highlights the superiority of our approach in real-world challenging scenarios where the generative models are unknown and the images may undergo various perturbations during dissemination. Another unusual observation is that increased perturbation magnitude correlates with higher AP scores in detection Diffusion model-generated images. One possible reason is that Diffusion images are less sensitive to external perturbations than real images thanks to the inherent denoising process of diffusion models \cite{luo2024lare}; this remains further exploration.

\subsection{Discussion}
The experimental results underscore the effectiveness of our proposed fake image detection approach. This method not only demonstrates robust performance across various generative models but also shows resilience against post-processing manipulations. These capabilities make it a powerful tool in the fight against the increasing spread of fake images.
Despite the unexpected trend observed in the charts, where our model's performance seems to improve as the image quality decreases and Sigma increases, we believe this could be attributed to the noise addition process inherent in the diffusion model's generation. This noise may render the generated images more distinguishable from authentic ones at lower qualities, thereby potentially enhancing our detection model's ability to identify them as fake.
Further investigation into the model's behavior under these conditions is necessary to fully understand and explain these results. It may involve examining the specific characteristics of the noise introduced by the diffusion model and how they interact with the detection model's features.

\section{CONCLUSIONS}
In conclusion, the rapid advancement of AI-synthesized image generation technologies presents a significant challenge to the authenticity and reliability of digital content. This work introduces a robust AI-Synthesized Image Detection framework based on Frequency-aware Multi-Feature Fusion, designed to counter the threat posed by these sophisticated forgery techniques. By integrating spatial, gradient, frequency domain, and CLIP features through a deep learning architecture, our model leverages a diverse set of visual representations to enhance detection accuracy. Employing an attention mechanism allows for dynamic learning of feature importance, optimizing detection performance. Extensive experiments on a comprehensive dataset demonstrate our method's exceptional accuracy, generalization, and robustness against various attack scenarios. As AI-synthesized image generation techniques evolve, our framework's adaptability and robustness make it a potent tool in the ongoing battle against the spread of fake images. This work contributes to the current state-of-the-art AI-synthesized image detection. It lays a solid foundation for future research to develop more generalized and reliable detection solutions.

\bibliographystyle{splncs04}
\bibliography{samplepaper}

\end{document}